\documentclass[12pt]{article}
\usepackage{epsfig}
\usepackage{amsmath}
\usepackage{amssymb}
\usepackage{amsfonts}
\usepackage{latexsym}
\usepackage{wasysym}
\usepackage{multirow}
\usepackage{fixmath}
\usepackage{color}
\usepackage{bbm}
\usepackage{ifsym}
\newcommand{\vn}{\boldsymbol{n}}
\definecolor{ForestGreen}{rgb}{0.133305, 0.545106, 0.133305}

\begin{document}

\title{\centering{\bf Nucleon axial charge and pion decay constant\\ from
  two-flavor lattice QCD\\}}

\author{R.~Horsley$^{1}$,
  Y.~Nakamura$^{2}$, A.~Nobile$^{3}$, 
  P.E.L.~Rakow$^{4}$,  \\
G.~Schierholz$^{5}$ and J.M. Zanotti$^{6}$\\[0.5em]
$^1$ School of Physics and Astronomy, University of Edinburgh,\\ Edinburgh
EH9 3JZ, United Kingdom\\ 
$^2$ RIKEN Advanced Institute for Computational Science,\\ Kobe, Hyogo
650-0047, Japan\\ 
$^3$ JSC, Forschungszentrum J\"ulich, 52425 J\"ulich, Germany\\ 
$^4$ Theoretical Physics Division, Department of Mathematical Sciences,\\
University of Liverpool, Liverpool L69 3BX, United Kingdom\\ 
$^5$ Deutsches Elektronen-Synchrotron DESY, 22603 Hamburg, Germany \\ 
$^6$ CSSM, School of Chemistry and Physics, University of Adelaide,\\
Adelaide SA 5005, Australia\\
\vspace*{-0.5cm}
}


\date{}
\maketitle

\begin{abstract}

The axial charge of the nucleon $g_A$ and the pion decay constant
$f_\pi$ are computed in two-flavor lattice QCD. The simulations are
carried out on lattices of various volumes and lattice
spacings. Results are reported for pion masses as low as
$m_\pi=130\,\mbox{MeV}$. Both quantities, $g_A$ and $f_\pi$, suffer from 
large finite size effects, which to leading order ChEFT and ChPT
turn out to be identical. By considering the naturally renormalized
ratio $g_A/f_\pi$, we observe a universal behavior as a function of
decreasing quark mass. From extrapolating the ratio to the physical point,
we find $g_A^R=1.29(5)(3)$, using the physical value of $f_\pi$ as
input and $r_0=0.50(1)$ to set the scale. In a
subsequent calculation we attempt to extrapolate $g_A$ and $f_\pi$
separately to the infinite volume. Both volume and quark mass
dependencies of $g_A$ and $f_\pi$ are found to be well decribed by
ChEFT and ChPT. We find at the physical point $g_A^R=1.24(4)$ and
$f_\pi^R=89.6(1.1)(1.8)\,\mbox{MeV}$. Both sets of results are in  
good agreement with  experiment. As a by-product we obtain the
low-energy constant $\bar{l}_4=4.2(1)$. 

\end{abstract}



\section{Introduction}

The axial charge $g_A$ of the nucleon is a fundamental measure of
nucleon structure. While $g_A$ has been known accurately for many
years from neutron $\beta$ decays, a calculation of $g_A$ from first
principles still
presents a significant challenge. Present lattice
calculations~\cite{Edwards:2005ym,Yamazaki:2008py,Alexandrou:2010hf,Capitani:2012gj,Lin:2011ab},
except perhaps~\cite{Capitani:2012gj},
underestimate the experimental value
by a large amount. The resolution of this problem is of great
importance to any further calculation of hadron structure.

\begin{table}[b!]
\begin{center}
\begin{tabular}{|c|c|c|c|c|c|c|}\hline
$\beta$ & $\kappa$ & Volume & $am_\pi$ & $g_A$ & $af_\pi$ &
  $r_0/a$\\ \hline
 $5.25$ & $0.13460$ & $16^3\times 32$ & $0.4932(10)$ & $1.442(13)\phantom{0}$ &
  $0.0886(8)\phantom{0}$ & \multirow{5}{*}{$6.603(53)$}\\ 
 $5.25$ & $0.13520$ & $16^3\times 32$ & $0.3821(13)$ & $1.438(20)\phantom{0}$ &
  $0.0756(8)\phantom{0}$ & \\ 
 $5.25$ & $0.13575$ & $24^3\times 48$ & $0.2556(5)\phantom{0}$  & $1.456(10)\phantom{0}$ &
  $0.0635(5)\phantom{0}$  & \\ 
 $5.25$ & $0.13600$ & $24^3\times 48$ & $0.1840(7)\phantom{0}$  & $1.412(18)\phantom{0}$ &
  $0.0550(4)\phantom{0}$  & \\ 
 $5.25$ & $0.13620$ & $32^3\times 64$ & $0.0997(11)$ & $1.368(51)\phantom{0}$ &
  $0.0439(6)\phantom{0}$  & \\ \hline
 $5.29$ & $0.13400$ & $16^3\times 32$ & $0.5767(11)$ & $1.437(12)\phantom{0}$ &
 $0.0936(9)\phantom{0}$  & \multirow{14}{*}{$7.004(54)$}\\ 
 $5.29$ & $0.13500$ & $16^3\times 32$ & $0.4206(9)\phantom{0}$  & $1.409(12)\phantom{0}$ &
 $0.0778(5)\phantom{0}$   & \\ 
 $5.29$ & $0.13550$ & $12^3\times 32$ & $0.3605(32)$ & $1.181(60)\phantom{0}$ &
 $0.0568(8)\phantom{0}$  & \\ 
 $5.29$ & $0.13550$ & $16^3\times 32$ & $0.3325(14)$ & $1.371(20)\phantom{0}$ &
 $0.0675(6)\phantom{0}$   & \\  
 $5.29$ & $0.13550$ & $24^3\times 48$ & $0.3270(6)\phantom{0}$  & $1.459(11)\phantom{0}$ & 
 $0.0689(7)\phantom{0}$  & \\ 
 $5.29$ & $0.13590$ & $12^3\times 32$ & $0.3369(62)$ & $0.967(105)$ &
 $0.0345(9)\phantom{0}$  & \\ 
 $5.29$ & $0.13590$ & $16^3\times 32$ & $0.2518(15)$ & $1.271(32)\phantom{0}$ &
 $0.0559(5)\phantom{0}$   & \\  
 $5.29$ & $0.13590$ & $24^3\times 48$ & $0.2395(5)\phantom{0}$  & $1.426(7)\phantom{00}$  & 
 $0.0588(3)\phantom{0}$   & \\ 
 $5.29$ & $0.13620$ & $24^3\times 48$ & $0.1552(6)\phantom{0}$  & $1.334(18)\phantom{0}$ &
 $0.0478(3)\phantom{0}$   & \\ 
 $5.29$ & $0.13632$ & $24^3\times 48$ & $0.1112(9)\phantom{0}$  & $1.271(67)\phantom{0}$ &
 $0.0398(4)\phantom{0}$   & \\ 
 $5.29$ & $0.13632$ & $32^3\times 64$ & $0.1070(5)\phantom{0}$  & $1.409(24)\phantom{0}$ &
 $0.0440(3)\phantom{0}$   & \\ 
 $5.29$ & $0.13632$ & $40^3\times 64$ & $0.1050(3)\phantom{0}$  & $1.439(17)\phantom{0}$ &
 $0.0445(3)\phantom{0}$   & \\ 
 $5.29$ & $0.13640$ & $40^3\times 64$ & $0.0660(8)\phantom{0}$  &
  $1.363(105)$ &  
 $0.0375(5)\phantom{0}$   & \\
 $5.29$ & $0.13640$ & $48^3\times 64$ & $0.0570(7)\phantom{0}$  & $1.572(52)\phantom{0}$ & 
 $0.0408(11)$  & \\ \hline
 $5.40$ & $0.13500$ & $24^3\times 48$ & $0.4030(4)\phantom{0}$  & $1.474(7)\phantom{00}$  & 
 $0.0691(5)\phantom{0}$   & \multirow{8}{*}{$8.285(74)$}\\  
 $5.40$ & $0.13560$ & $24^3\times 48$ & $0.3123(7)\phantom{0}$  & $1.451(11)\phantom{0}$ & 
 $0.0620(5)\phantom{0}$   & \\  
 $5.40$ & $0.13610$ & $24^3\times 48$ & $0.2208(7)\phantom{0}$  & $1.410(20)\phantom{0}$ & 
 $0.0513(4)\phantom{0}$   & \\  
 $5.40$ & $0.13625$ & $24^3\times 48$ & $0.1902(6)\phantom{0}$  & $1.377(20)\phantom{0}$ & 
 $0.0470(3)\phantom{0}$   & \\  
 $5.40$ & $0.13640$ & $24^3\times 48$ & $0.1538(10)$ & $1.261(34)\phantom{0}$ &
 $0.0419(4)\phantom{0}$   & \\ 
 $5.40$ & $0.13640$ & $32^3\times 64$ & $0.1505(5)\phantom{0}$  & $1.402(17)\phantom{0}$ & 
 $0.0442(4)\phantom{0}$   & \\  
 $5.40$ & $0.13660$ & $32^3\times 64$ & $0.0845(6)\phantom{0}$  & $1.206(79)\phantom{0}$ &
 $0.0342(4)\phantom{0}$   & \\  
 $5.40$ & $0.13660$ & $48^3\times 64$ & $0.0797(3)\phantom{0}$  & $1.403(29)\phantom{0}$ & 
 $0.0362(3)\phantom{0}$   & \\ \hline 
\end{tabular}
\end{center}
\caption{Parameters of our lattice data sets, together with the pion
  mass, the bare axial charge and the pion decay constant. Also listed
are the chirally extrapolated values of $r_0/a$. In this work we use
$r_0=0.50(1)\,\mbox{fm}$ to convert lattice numbers to physical units.}  
\label{tab1}
\end{table}

Lattice calculations of $g_A$ are in many ways connected to
calculations of the pion decay constant $f_\pi$. Both quantities
involve the axial vector current, which is not conserved and thus
needs to be renormalized. Though it is standard practice nowadays 
to compute the renormalization constant
nonperturbatively (see, for
example,~\cite{Martinelli:1994ty,Gockeler:2010yr}), some 
scope of uncertainty remains~\cite{Constantinou:2012dt}.
Another common 
feature is that $g_A$ and $f_\pi$ seem to be affected by large
finite size corrections, in particular at small pion masses, which to
leading  order ChEFT and ChPT~\cite{Beane:2004rf,Khan:2006de,Colangelo:2005gd}
appear to be 
the same in both cases. This led us to suggest to determine $g_A$
from the ratio $g_A/f_\pi$. Preliminary results~\cite{Pleiter:2011gw}
were encouraging, and we present here the final analysis of this
investigation.  

The calculations are done with two flavors of nonperturbatively $O(a)$
improved Wilson fermions and Wilson plaquette
action~\cite{Bali:2012qs}, including simulations at virtually physical
pion mass and on a variety of lattice volumes. This allows for a separate
extrapolation of both $g_A$ and $f_\pi$ to the infinite
volume and the physical point.

\begin{figure}[b!]
\vspace*{-1.5cm}
\begin{center}
\epsfig{file=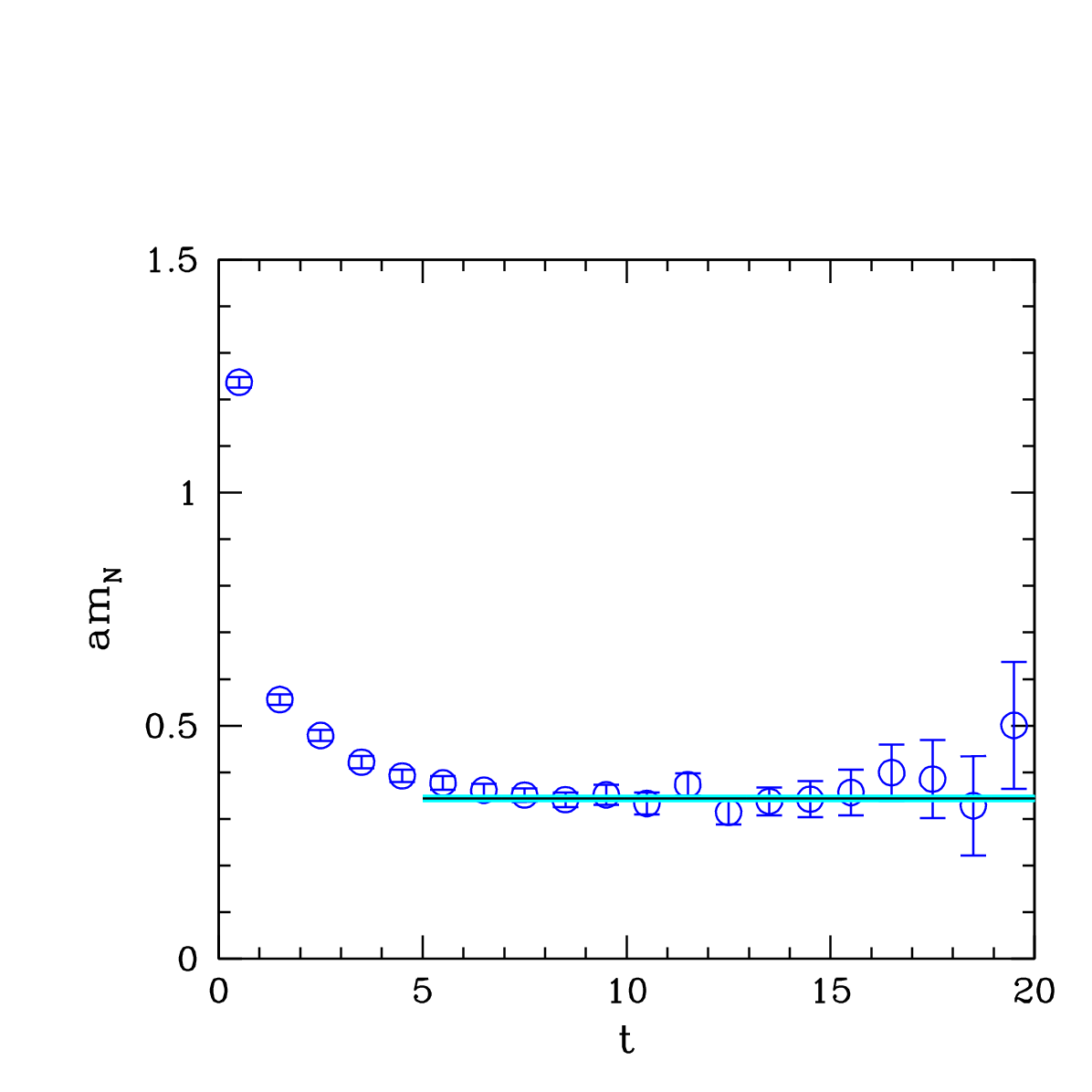,width=10cm,clip=}
\end{center}
\vspace*{-0.5cm}
\caption{The effective mass $m_N$ of the nucleon on the
  $48\times 64$ lattice at $\beta=5.29$, $\kappa=0.13640$,
  corresponding to our smallest pion mass. The horizontal line shows
  the fit and error band. The fit range for this nucleon mass was $t=8-16$.}   
\label{fignuc}
\end{figure}

\section{Lattice simulation}

Our lattice ensembles are listed in Table~\ref{tab1}. The pion masses
and the chirally extrapolated values of $r_0/a$ are taken from our
preceding paper~\cite{Bali:2012qs} on the nucleon mass and sigma
term. The Sommer parameter was found to be $r_0=0.50(1)\,\mbox{fm}$,
which we will use to set the scale throughout this paper. The 
ensembles cover three $\beta$ values, $\beta=5.25$, $5.29$ and
$5.40$, with lattice spacings $a=0.076$, $0.071$ and
$0.060\,\mbox{fm}$.

We employ the improved axial vector current
\begin{equation}
\mathcal{A}_\mu(x) = \bar{q}(x)\gamma_\mu\gamma_5q(x) + a c_A
\partial_\mu \bar{q}(x)\gamma_5q(x) \,,
\label{imp}
\end{equation}
where $c_A$ is taken from~\cite{Della Morte:2005se}. The improvement
term does not contribute to forward matrix elements, but it will
contribute to $f_\pi$. The calculation of $g_A$
follows~\cite{Gockeler:1995wg,Capitani:1998ff,Khan:2006de} with one
exception, namely that on the $48^3\times 64$ lattice at $\beta=5.29$,
$\kappa=0.13640$ we have employed Wuppertal smearing instead of
Jacobi smearing. It involves computing the ratio of two- and
three-point functions
\begin{equation}
R_{\alpha\beta}(t,\tau)
        =    \frac{\langle
        N_\alpha(t) \mathcal{A}_\mu(\tau) 
        \bar{N}_\beta(0) \rangle}
        {\langle N(t) \bar{N}(0) \rangle} \,,
\end{equation}
$\bar{N}$ and $N$ being the nucleon creation and annihilation
operators at zero momentum with Dirac indices $\alpha$, $\beta$, which
are used to project onto the appropriate nucleon spin. The spins of
the nucleon appearing in the denominator are summed over.
Any smearing of the source (at time $0$) and sink operators (at time
$t$) is cancelled in this ratio. For $\beta=5.4$ we use $t=17$, while
the lightest two ensembles at $\beta=5.29$, $\kappa=0.13640$ and
$\kappa=0.13632$, use $t=15$. All other ensembles use $t=13$. In physical
units this amounts to time separations between source and sink of
$\approx 1.1 \,\mbox{fm}$ at the smaller pion masses, which is current
state of the art (see Table 3 of~\cite{Owen:2012ts}). In
Fig.~\ref{fignuc} we plot the 
effective mass $m_N$ of the nucleon at our smallest, nearly
physical pion mass, which indicates that excited states have 
died out at times $t \gtrsim 5$. To determine the nucleon mass on this
ensemble, we chose the conservative fit range $t=8-16$.

\begin{figure}[t!]
\vspace*{-1.0cm}
\begin{center}
\epsfig{file=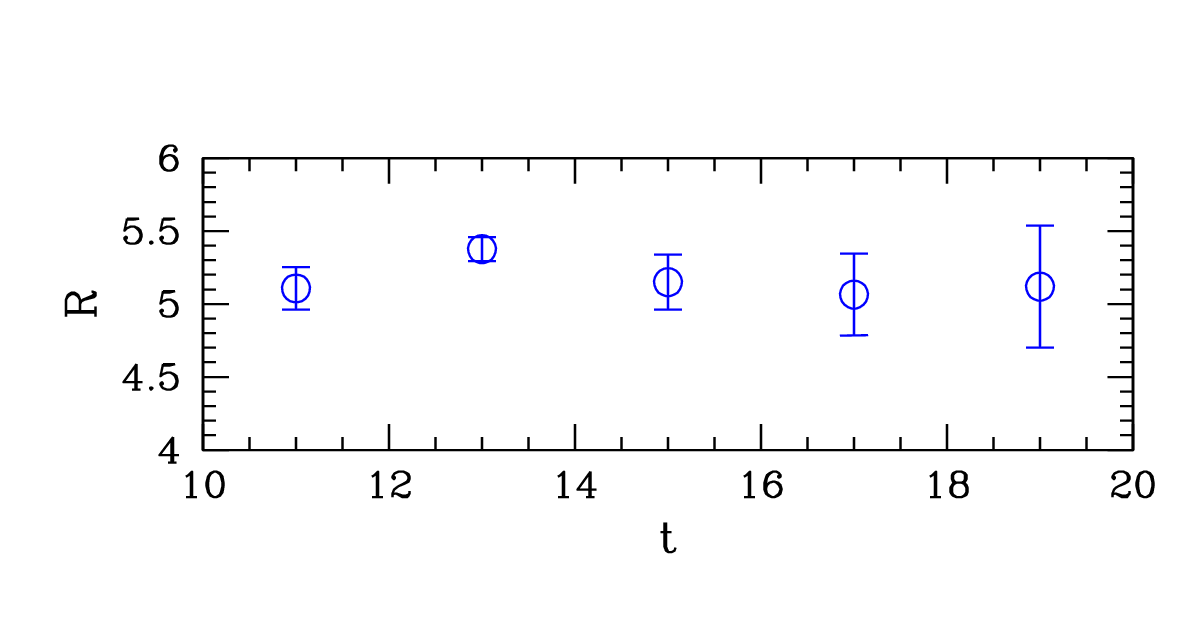,width=10cm,clip=}
\end{center}
\vspace*{-1.0cm}
\caption{The ratio R as a function of the source-sink time separation
  $t$ on the   $24^3\times 48$ lattice at $\beta=5.29$, $\kappa=0.13590$.} 
\label{fig1}
\end{figure}

In~\cite{Capitani:2012gj} it has been argued that contributions from
excited states might be the reason for lattice calculations to
underestimate $g_A$, when compared to its experimental value. To
investigate this scenario (beyond tuning the 
smearing parameters, see Fig.~\ref{fignuc}), we have performed 
additional simulations on the $24^3\times 48$
lattice at $\beta=5.29$, $\kappa=0.13590$ with a large range of
different source-sink separations, $t=11, \cdots, 19$ ($0.79, \cdots,
1.36\, \mbox{fm}$), albeit with
somewhat lower statistics than our reference point at $t=13$ ($0.93\,
\mbox{fm}$) on this ensemble. In 
Fig.~\ref{fig1} we show the ratio $R$ for various time 
separations $t$ between source and sink. If our $g_A$ determinations
were affected by excited state contaminations, then we should find a
larger value at separations $t>13$. However, we do not see any systematic
deviation of $R$ 
from our result at $t=13$ within the error bars, not even for
$t=11$. This provides us with confidence that our choices of $t$ are
sufficient with our choice of source and sink smearing. Similar
conclusions were found in~\cite{Alexandrou:2010hf}. 

Our smearing parameters are tuned to give a {\it rms} radius of
$\approx 0.5\,\mbox{fm}$, which is about half the radius of the
nucleon. For this level of smearing no further improvement of the
extracted result for $g_A$ was found by employing variational
techniques~\cite{Owen:2012ts}, which systematically separate excited
states out from the ground state at source and sink. 

The calculation of $f_\pi$ follows~\cite{Gockeler:2005mh}. 
We use the notation employed in ChPT, with the experimental value
$f_{\pi^+}=92.2\,\mbox{MeV}$. Our final results for the bare
quantities, $g_A$ and $af_\pi$, on all our ensembles, are given in
Table~\ref{tab1}. 

\section{Results}

Except for the very lowest pion mass, $m_\pi L \gtrsim 4$ ($L$ being
the spatial extent of the lattice) on our larger lattices at any 
other $\kappa$ value, which is state of the art for pion masses of 
$O(200)\,\mbox{MeV}$. But even on lattices of this size $g_A$, $f_\pi$
and $m_\pi$ are found to suffer from finite size 
effects, which we have to deal with in one way or another. 

Finite size corrections to $g_A$, $f_\pi$ and $m_\pi$ have been
studied extensively in the literature. In the Appendix we show, based
on predictions of ChEFT and ChPT adapted to the finite volume, that 
the leading corrections to $g_A$ and $f_\pi$ are identical and cancel
in the ratio $g_A/f_\pi$. This makes $g_A/f_\pi$ the preferred 
quantity for computing $g_A$. 

\subsection{The axial coupling $\boldsymbol{g_A}$ from the ratio
  $\boldsymbol{g_A(L)/f_\pi(L)}$} 

\begin{figure}[t!]
\vspace*{-1.05cm}
\begin{center}
\epsfig{file=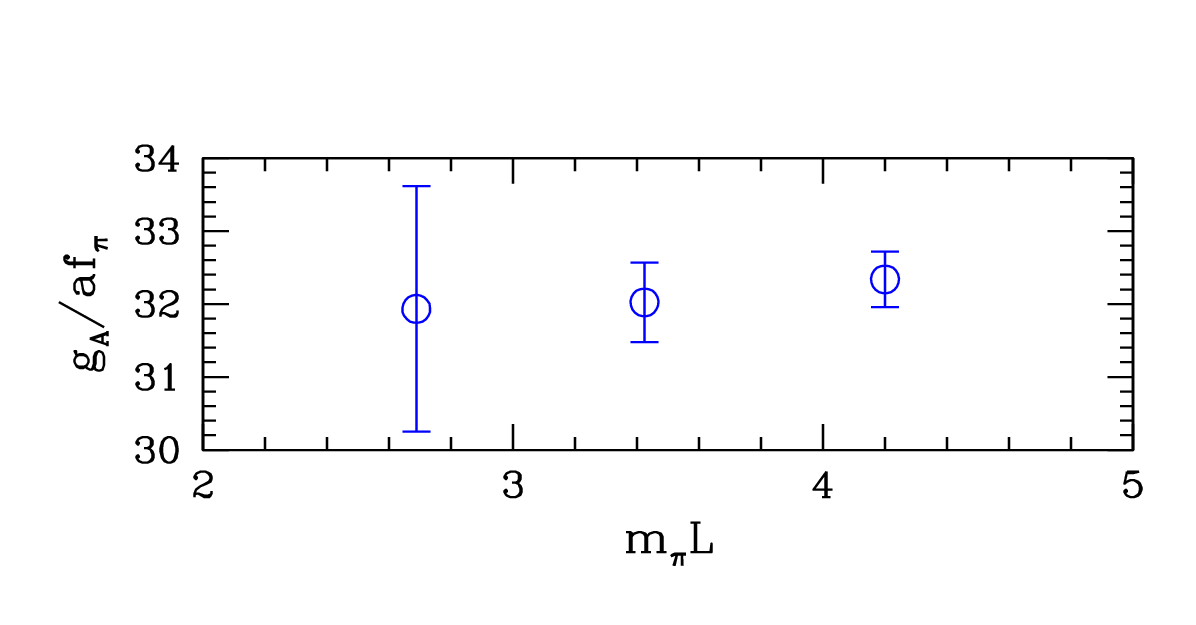,width=10cm,clip=}
\end{center}
\vspace*{-1.0cm}
\caption{The ratio $g_A(L)/af_\pi(L)$ as a function of $m_\pi L$ at
  $\beta=5.29$, $\kappa=0.13632$.} 
\label{figL}
\end{figure}

Neglecting 
NNLO and $O(\Delta(L))$ corrections, we obtain from eqs.~(\ref{fsga3})
and (\ref{fsfpi}) 
\begin{equation}
\frac{g_A(L)-g_A(\infty)}{g_A(\infty)} =
\frac{f_\pi(L)-f_\pi(\infty)}{f_\pi(\infty)} \,.
\label{fsgafpi}
\end{equation}
Denoting the physical, renormalized axial charge and pion decay
constant in the infinite volume by $g_A^R$ and $f_\pi^R$,
respectively, and making use of the fact that the renormalization
constant $Z_A$ of the axial vector current cancels in the ratio
$g_A/f_\pi$, we then have 
\begin{equation}
\frac{g_A^R}{f_\pi^R} = \frac{g_A(\infty)}{f_\pi(\infty)} =
\frac{g_A(L)}{f_\pi(L)} \,. 
\label{ratinf}
\end{equation}
To test this relation, we plot $g_A/af_\pi$ for three different
lattice volumes at our second lowest pion mass in Fig.~\ref{figL}. The
ratio is found to be independent of the volume, within the errors,
which demonstates that finite size corrections cancel indeed in
$g_A/f_\pi$. 

\begin{figure}[t!]
\vspace*{-2.0cm}
\begin{center}
\epsfig{file=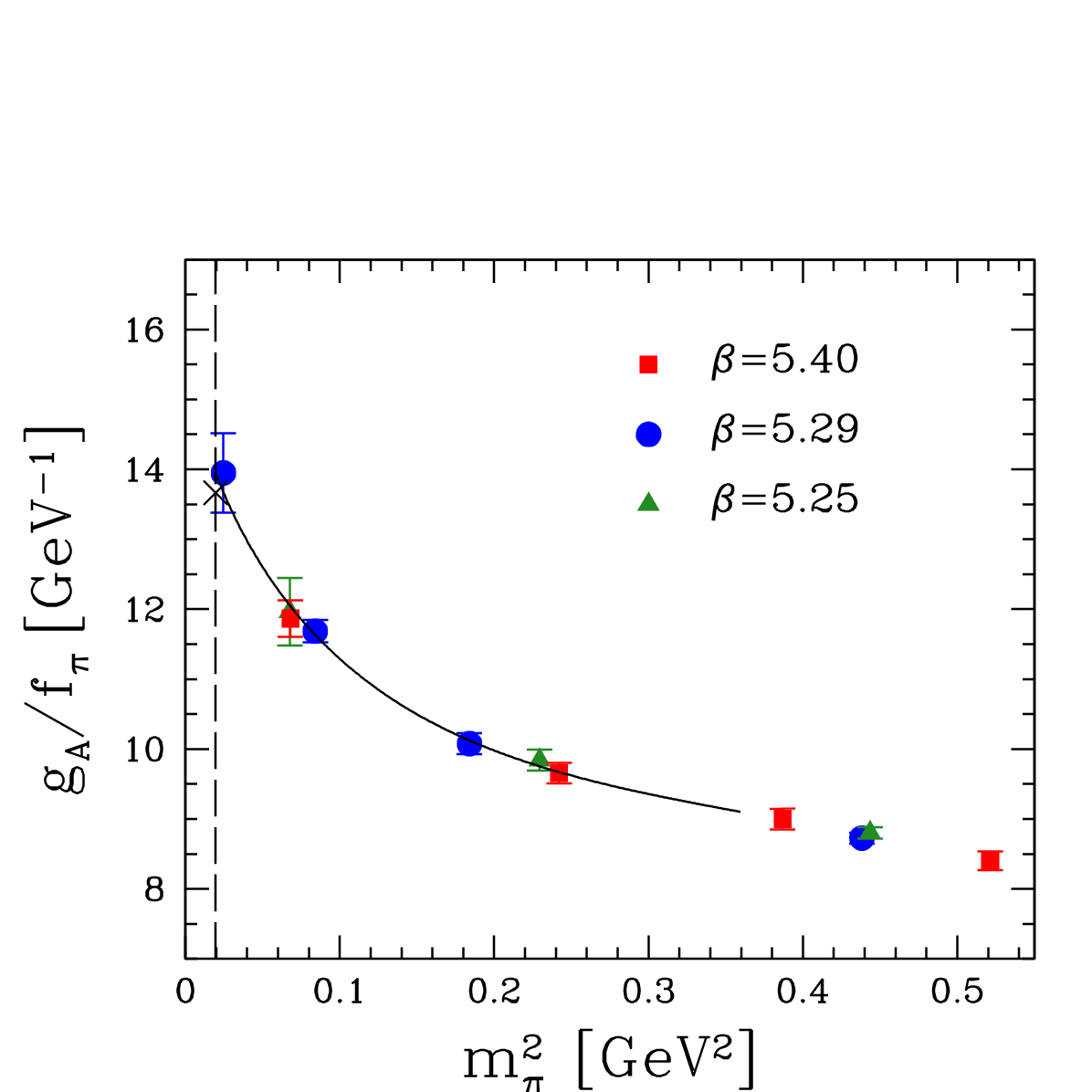,width=10cm,clip=}
\end{center}
\vspace*{-0.5cm}
\caption{The ratio $g_A(L)/f_\pi(L)$ as a function of $m_\pi^2(L)$,
  together with the experimental value 
  ($\times$). The curve shows a fit of eq.~(\ref{chpt}) to the data.}   
\label{fig6}
\end{figure}

Let us now turn to the calculation of $g_A$. In Fig.~\ref{fig6} we
plot the ratio $g_A(L)/f_\pi(L)$ for our (raw) 
data points listed in Table~\ref{tab1}, restricting ourselves to pion
masses $m_\pi \leq 750\,\mbox{MeV}$, and taking
$r_0=0.50(1)\,\mbox{fm}$ to set the scale. If we have more than one 
volume at a given $\kappa$ value, we show the result of the largest
volume. The lowest pion mass in Fig.~\ref{fig6} is
$157\,\mbox{MeV}$. The data points of all three $\beta$ values lie nicely on a
universal curve. At $m_\pi^2 \approx 0.06, \,
0.23$ and $0.44 \,\mbox{GeV}^2$, for example, where we have results
for more than one lattice spacing, the data points coincide with each
other, indicating that discretization effects are negligible.  

With finite size corrections being practically absent,
the leading order chiral expansion of $g_A/f_\pi$ 
can be cast in the form~\cite{Procura:2006gq,Khan:2006de,Colangelo:2001df} 
\begin{equation}
\frac{g_A}{f_\pi} = A + B\, m_\pi^2 + C\, m_\pi^2\, \ln{m_\pi^2} +D\,
m_\pi^4  \,.
\label{chpt}
\end{equation}
We have fitted eq.~(\ref{chpt}) to the data points in
Fig.~\ref{fig6}. The result is shown by the solid curve. At the
physical point this gives 
\begin{equation}
\frac{g_A}{f_\pi}=13.95\pm 0.71 \pm 0.30\; \mbox{GeV}^{-1} \,.
\label{rat}
\end{equation}
The second error is due to the error on $r_0$. 
Multiplying the ratio (\ref{rat}) by the physical value of $f_\pi$,
$f_\pi^R=92.2\,\mbox{MeV}$, we then obtain
\begin{equation}
g_A^R=1.29 \pm 0.05 \pm 0.03\,.
\end{equation}
Alternatively, we could have set the scale by
the physical value of $f_\pi$, using the results of
Sec.~\ref{axinf}. That would give the value $g_A^R=1.27(5)$.

Now that we have presented the main result of the paper, i.e.\ $g_A^R$ from
the ratio $g_A(L)/f_\pi(L)$, we 
proceed to study $g_A^R$ and $f_\pi^R$ separately and present results in
the context of established expressions from finite volume ChEFT and
ChPT. 

\subsection{$\boldsymbol{g_A}$ and $\boldsymbol{f_\pi}$ in the
  infinite volume}    
\label{axinf}

We now consider explicitly the finite size formulae as given in the
Appendix. In the following fits we take
$f_0=86\,\mbox{MeV}$~\cite{Colangelo:2003hf}. There is some freedom in
which pion mass to take in 
eqs.~(\ref{fsga3}), (\ref{fsfpi}) and (\ref{com}). We choose
$m_\pi=m_\pi(\infty)$ 
in $\lambda$, $\lambda(y)$ and $c(m_\pi)$, and $m_\pi=m_\pi(L)$ 
otherwise.  


\begin{figure}[t!]
\vspace*{-1.5cm}
\begin{center}
\epsfig{file=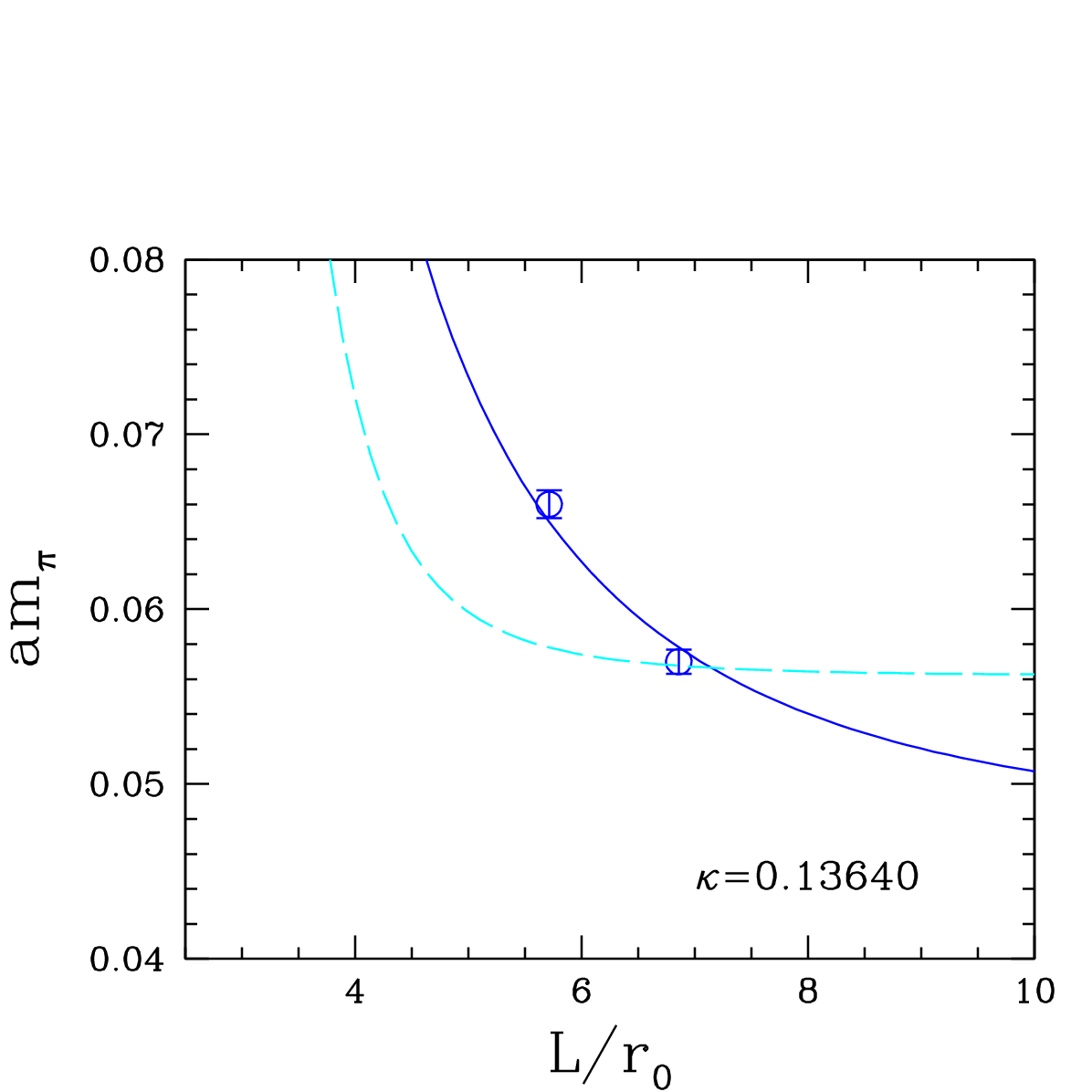,width=8cm,clip=}
\epsfig{file=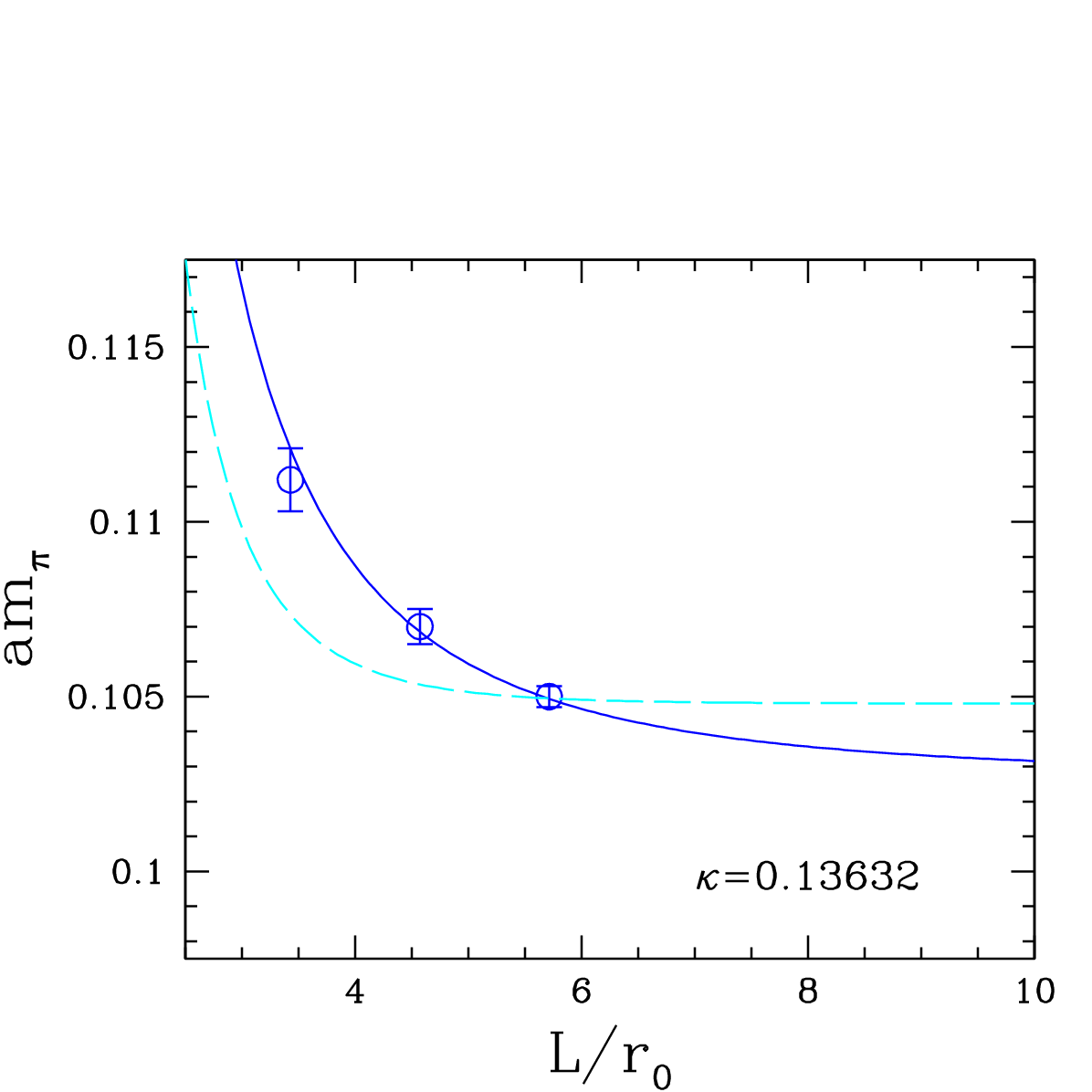,width=8cm,clip=}
\end{center}
\vspace*{-0.5cm}
\caption{The pion mass $am_\pi$ as a function of lattice size for two
  ensembles at $\beta=5.29$. The solid line shows a fit of eq.~(\ref{com})
  to the data. The dashed line shows the NLO result, eq.~(\ref{fsmpi}),
  fitted to the smallest mass point.} 
\label{fig2}
\end{figure}

Let us first consider the pion mass. In Fig.~\ref{fig2} we show the
fits of eq.~(\ref{com}) to $m_\pi$ for two of our lattice
ensembles. The corrections to 
$m_\pi$ are well described by this equation. Apart from
$m_\pi(\infty)$, we have one free parameter, $c(m_\pi)$, only. Equally 
good fits are obtained for $\beta=5.40$, 
$\kappa=0.13660$ and $0.13640$. The parameter $c(m_\pi)$ is found to
vanish with a large inverse power of the pion mass.\footnote{At
  $\beta=5.29$, $\kappa=0.13632$, i.e. our second smallest pion mass,
  $c(m_\pi)$ has dropped  
  to the value $0.15$ already.} The finite size
corrections predicted by the NLO expression
(\ref{fsmpi}), on the other hand, are nowhere near as big as the
effect shown by the data. 
In Table~\ref{tab2} we list our final pion masses. Our lowest mass
turns out to be $m_\pi=130(5)\,\mbox{MeV}$. 

\begin{table}[t!]
\begin{center}
\vspace*{0.25cm}
\begin{tabular}{|c|c|c|c|c|}\hline
$\beta$ & $\kappa$ & $m_\pi\,[\mbox{MeV}]$ & $g_A^R$ &
  $f_\pi^R\,[\mbox{MeV}]$ \\ \hline
 $5.25$ & $0.13600$ & $479(2)$ & $1.07(1)$  & $108.9(0.8)$ \\ 
 $5.29$ & $0.13620$ & $426(2)$ & $1.05(2)$  & $103.6(0.6)$ \\ 
 $5.29$ & $0.13632$ & $284(2)$ & $1.10(2)$  & $94.7(0.6)\phantom{0}$ \\ 
 $5.29$ & $0.13640$ & $130(5)$ & $1.24(4)$  & $89.7(1.5)\phantom{0}$ \\ 
 $5.40$ & $0.13640$ & $492(2)$ & $1.09(1)$  & $112.3(0.9)$ \\  
 $5.40$ & $0.13660$ & $253(2)$ & $1.09(2)$  & $93.0(0.7)\phantom{0}$ \\ \hline 
\end{tabular}
\end{center}
\vspace*{-0.25cm}
\caption{The pion  mass and the renormalized axial coupling and pion
  decay constant extrapolated to the infinite volume for $m_\pi \leq
  500\,\mbox{MeV}$ and $r_0=0.50\,\mbox{fm}$.}   
\label{tab2}
\end{table}

\begin{figure}[t!]
\vspace*{-1cm}
\begin{center}
\epsfig{file=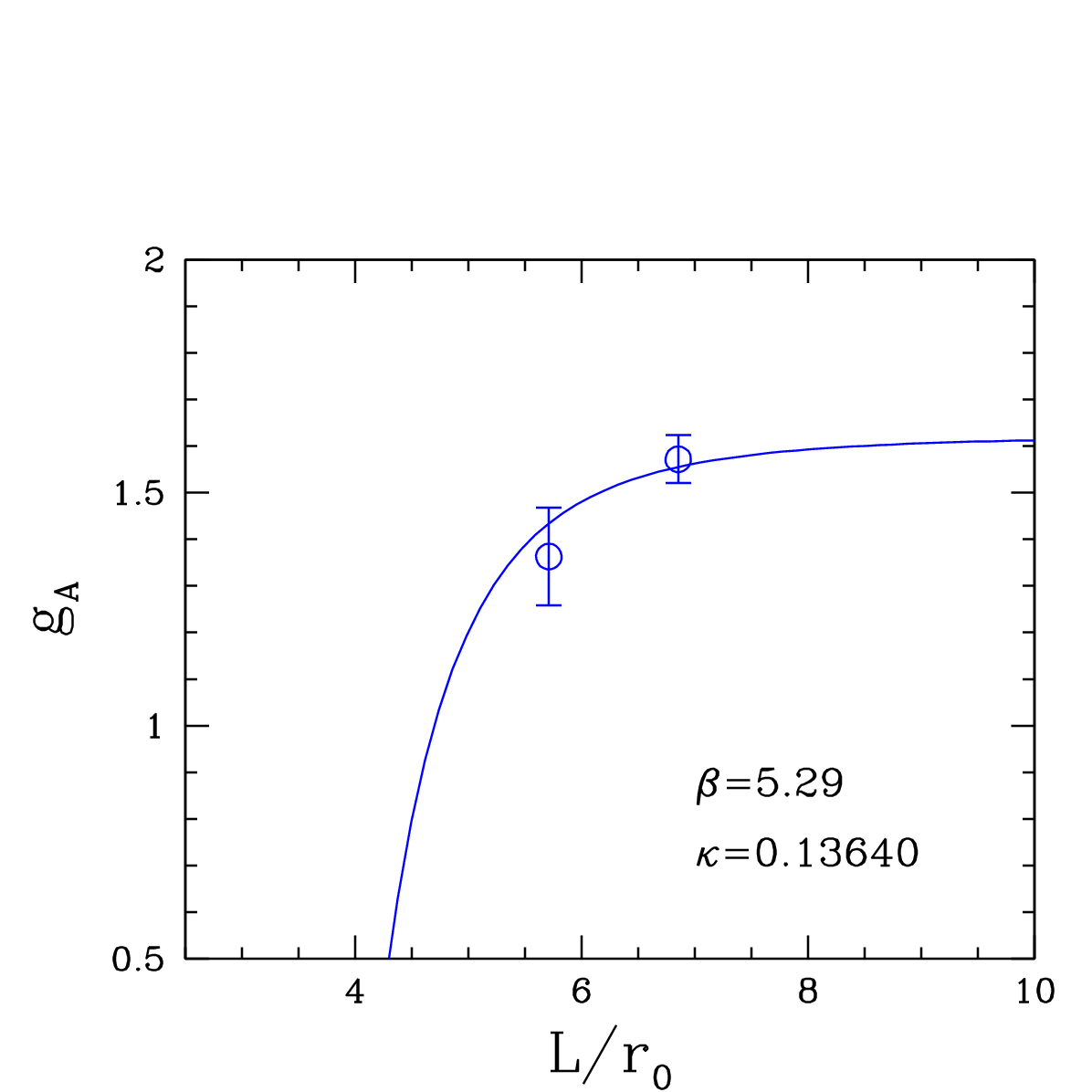,width=8cm,clip=}
\epsfig{file=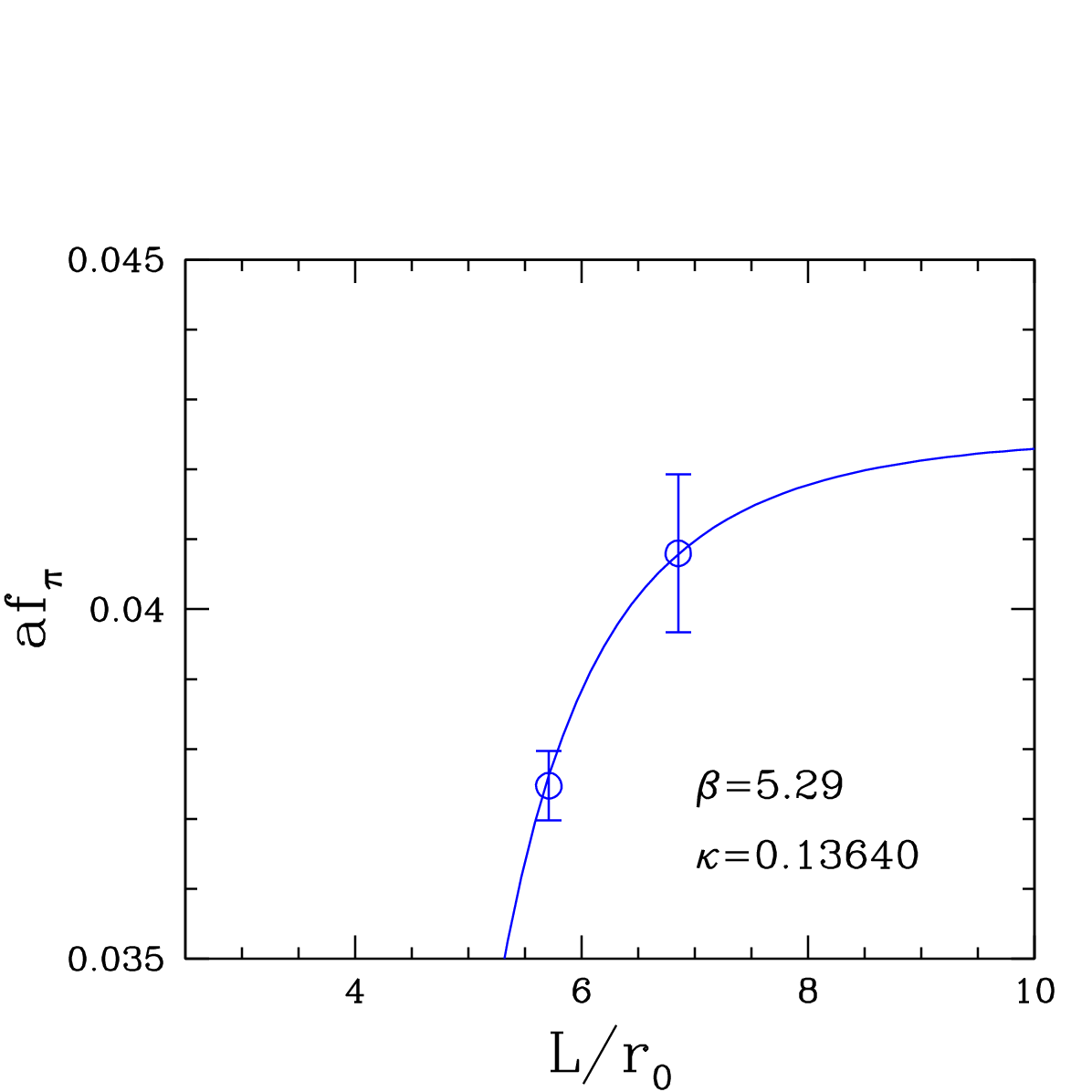,width=8cm,clip=}
\\[-3.0em]
\epsfig{file=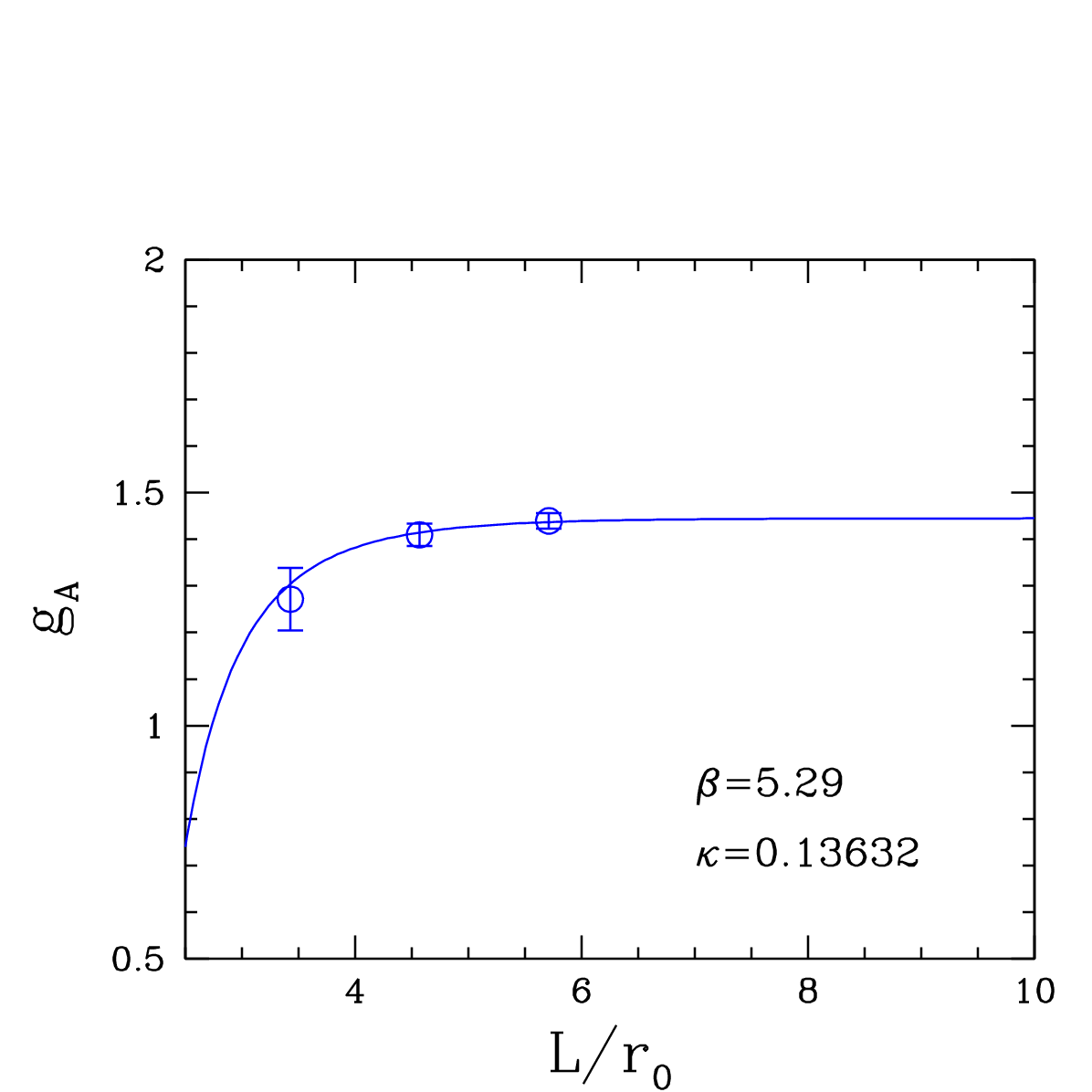,width=8cm,clip=}
\epsfig{file=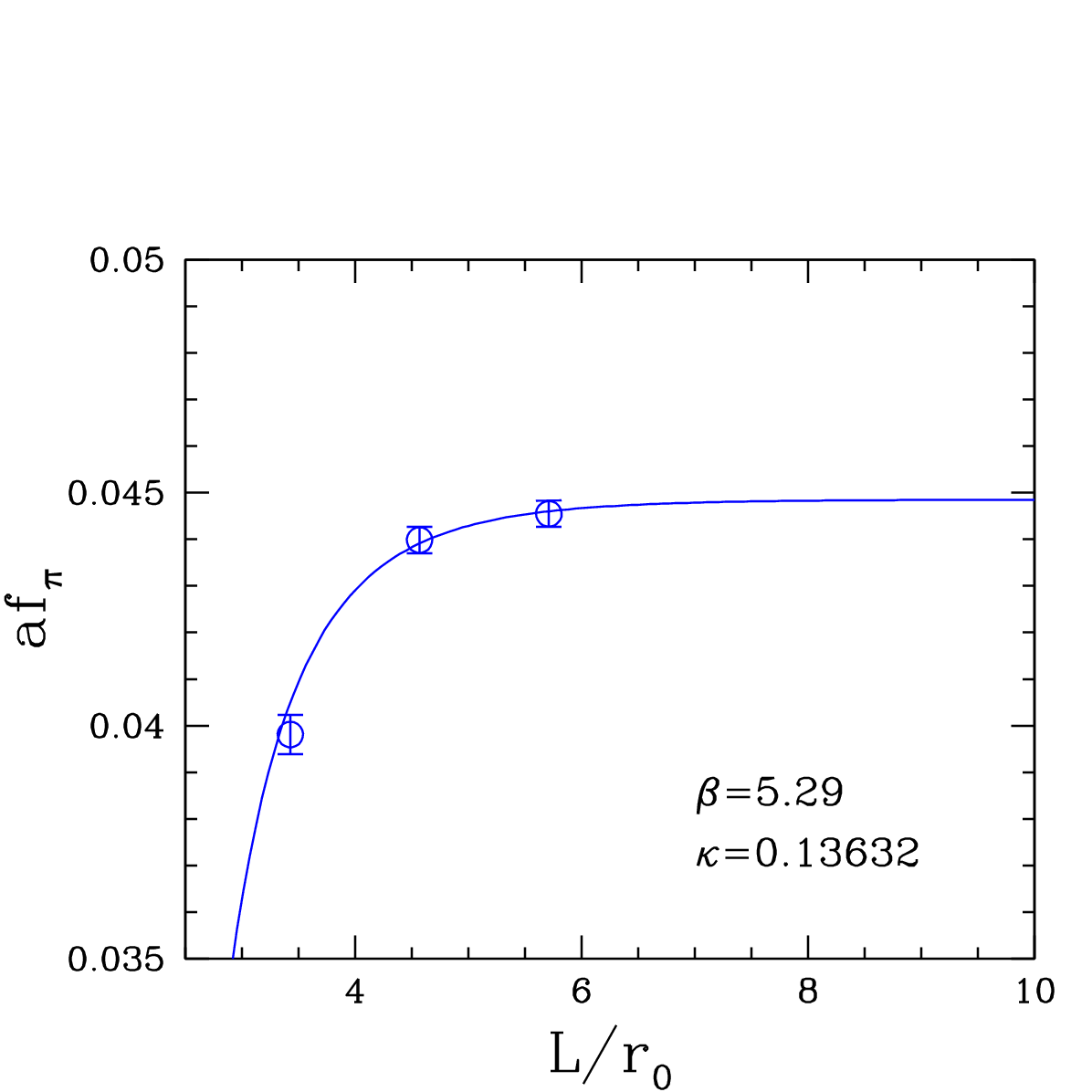,width=8cm,clip=}
\\[-3.0em]
\epsfig{file=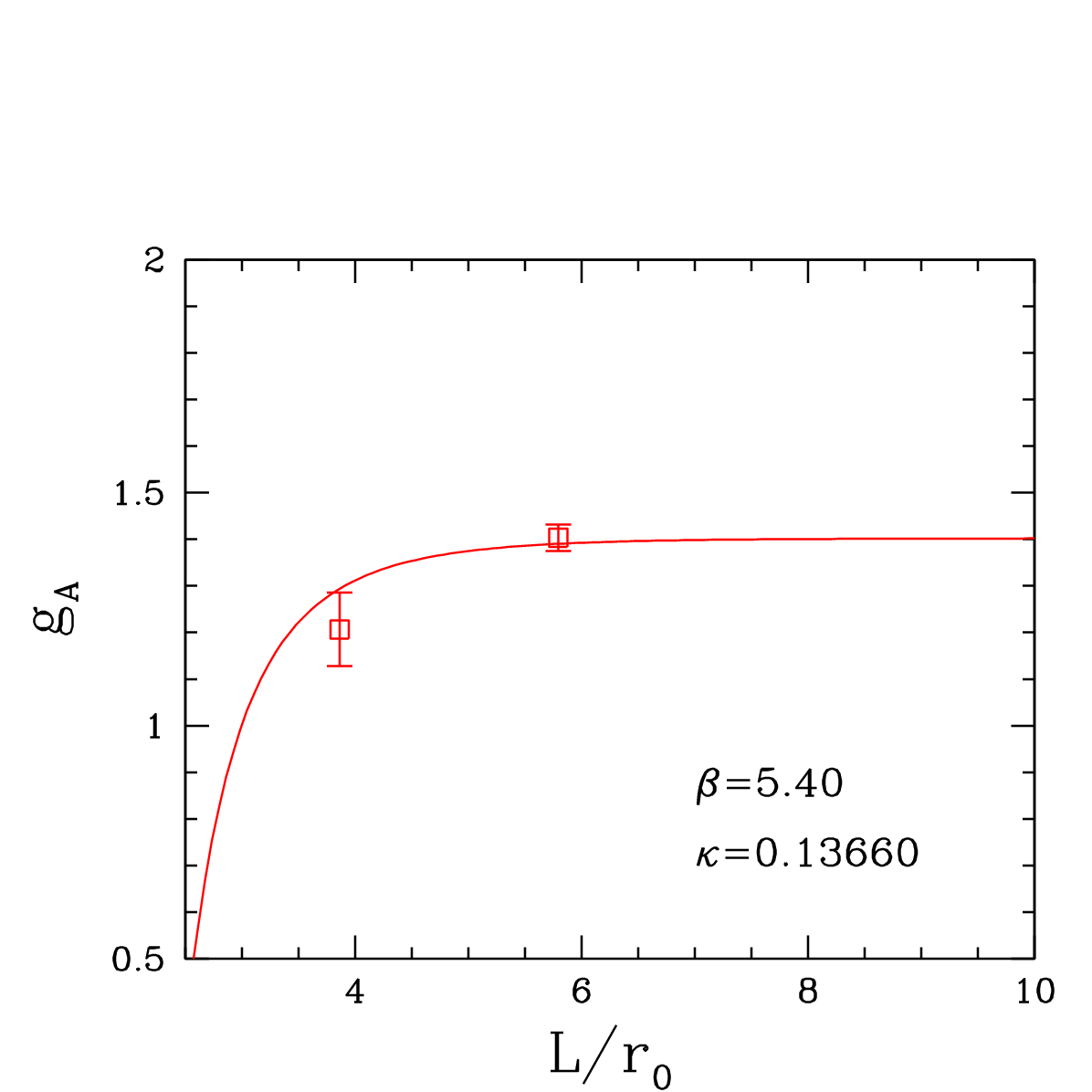,width=8cm,clip=}
\epsfig{file=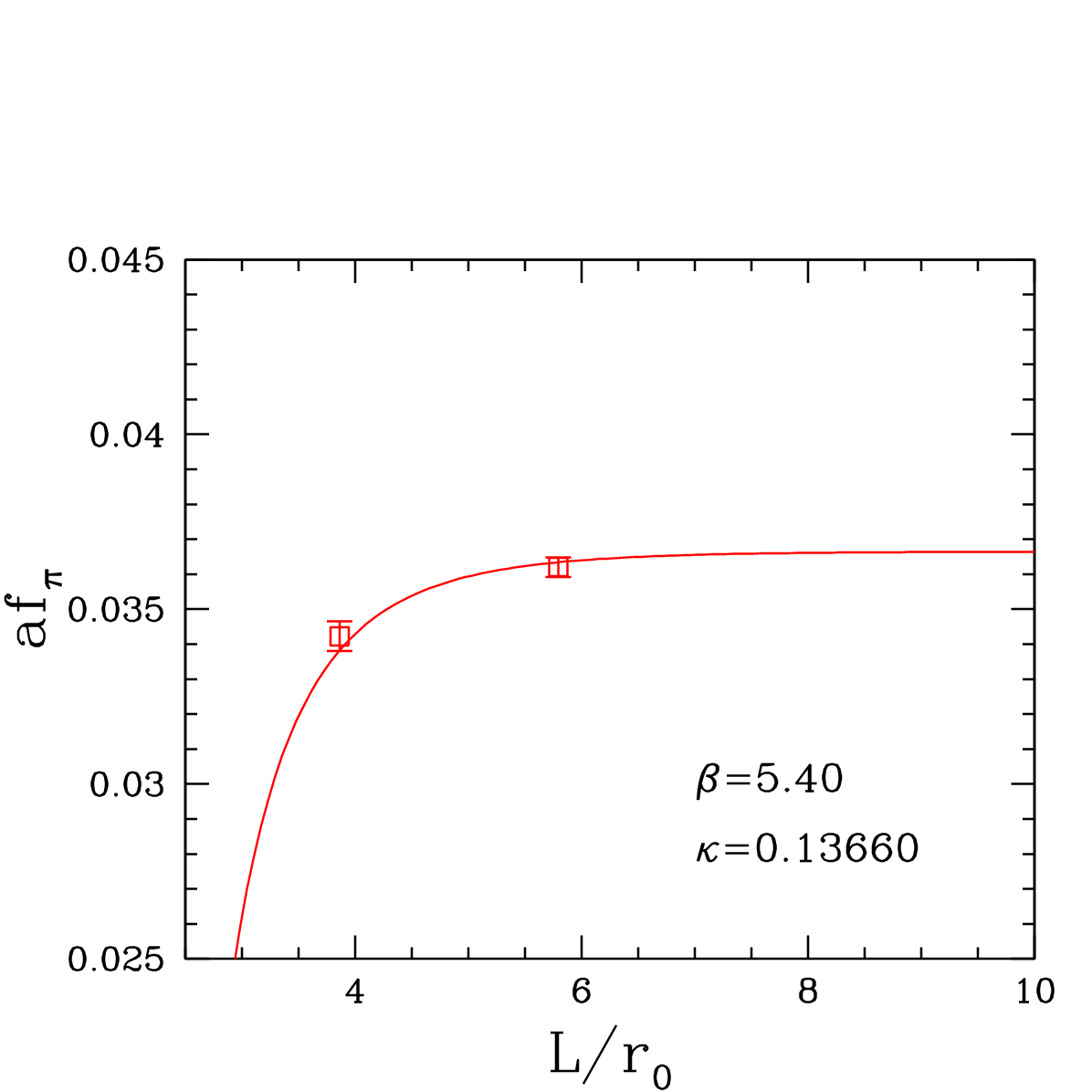,width=8cm,clip=}
\\[-2.5em] 
\end{center}
\caption{The bare axial charge $g_A$ and the bare pion decay constant
  $af_\pi$ as a function of the spatial extent of the lattice, together
  with the leading order finite size corrections of eqs.~(\ref{fsga3})
  and (\ref{fsfpi}).}
\label{fig3}
\end{figure}

Let us now turn to the axial charge and the pion decay constant. In
Fig.~\ref{fig3} we show the fits of eqs.~(\ref{fsga3}) and
(\ref{fsfpi}) to $g_A$ and $af_\pi$, respectively, for our three
lowest pion masses. In this case the fits involve one free parameter each,
$g_A(\infty)$ and $af_\pi(\infty)$, only. The leading order expressions are
able to describe the data at $\beta = 5.29$, $\kappa=0.13632$ on all
three volumes, which include data with $m_\pi L < 3$ as well as $m_\pi
L > 4$. This gives us confidence that the fits 
provide a reasonable
infinite volume extrapolation at the lighter mass point as well, where
we do not have access to data with larger $m_\pi L$.\footnote{It
  should be noted though that $m_\pi L$ is not the ultimate benchmark,
  contrary to common belief. With decreasing pion mass the
  corrections turn into a $1/L^3$ behavior. See
  also~\cite{Bietenholz:2010az}.}   
All fits gave $\chi^2/\mathrm{d.o.f} < 1.4$.

To obtain continuum numbers, we need to renormalize the axial vector
current. The latter reads
\begin{equation}
\mathcal{A}_\mu^R  = Z_A \left(1+b_A am_q \right)\, \mathcal{A}_\mu \,.
\end{equation}
The coefficient $b_A$ is required to maintain $O(a)$ improvement for
nonvanishing quark masses $m_q$ as well. The renormalization constant $Z_A$
has been computed nonperturbatively in~\cite{Gockeler:2010yr},
employing the Rome-Southampton method~\cite{Martinelli:1994ty}, with
the result
\begin{equation}
\begin{tabular}{c|ccc}
$\beta$ & $5.25$ & $5.29$ & $5.40$ \\ \hline
$Z_A$ & $0.760(1)$ & $0.764(1)$ & $0.777(1)$\\
\end{tabular}
\vspace*{0.25cm}
\end{equation}

\clearpage

\clearpage

\begin{figure}[t!]
\vspace*{-1.75cm}
\begin{center}
\epsfig{file=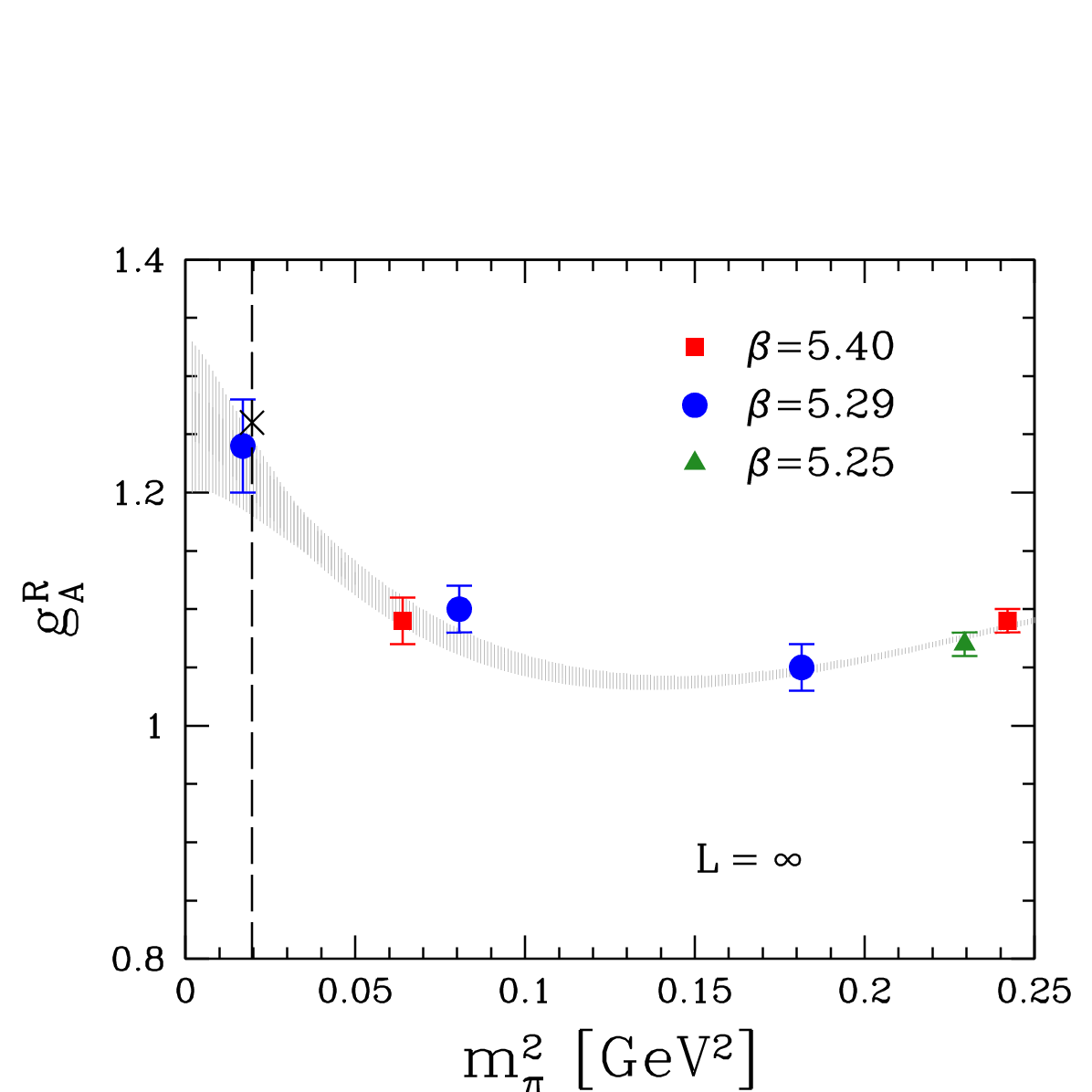,width=10cm,clip=}
\end{center}
\caption{The renormalized axial charge $g_A^R$ in the infinite volume
plotted against $m_\pi^2(\infty)$, together with the experimental
value $g_A=1.27$ ($\times$). The shaded area shows the fit of
eq.~(\ref{cheft}) to the data.}   
\label{fig4}
\vspace*{-1.0cm}
\begin{center}
\epsfig{file=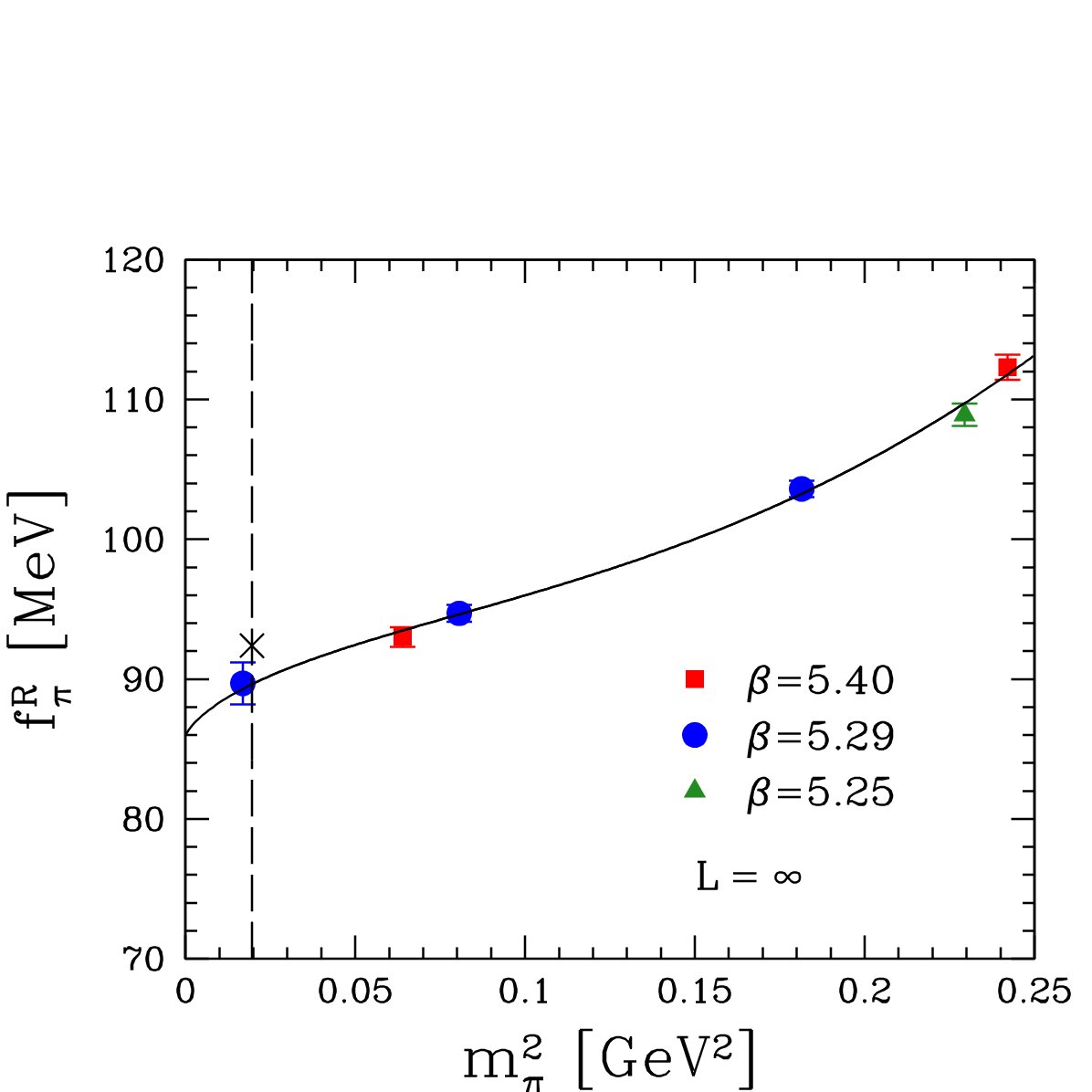,width=10cm,clip=}
\end{center}
\caption{The renormalized pion decay constant $f_\pi^R$ in the
infinite volume plotted against $m_\pi^2(\infty)$, together with
the experimental value $f_\pi=92.2\,\mbox{MeV}$ ($\times$). The curve
shows a fit of 
eq.~(\ref{chpt2}) to the data.}   
\label{fig5}
\end{figure}

\clearpage

\noindent
The coefficient $b_A$ is only known perturbatively~\cite{Sint:1997jx},
\begin{equation}
b_A=1+0.1522\,g^2\,.
\end{equation}

In Table~\ref{tab2} we give $g_A^R$ in the infinite volume. For pion
masses $m_\pi \leq 300 \,\mbox{MeV}$ we demand
that we have at least two lattice volumes to ensure a controlled
extrapolation. For this reason we excluded the point at 
$\beta=5.25$, $\kappa=0.13620$ from the analysis.  

Our results for $g_A^R$ are plotted in Fig.~\ref{fig4}. Since after
finite volume corrections the lightest pion mass is $130\,\mbox{MeV}$,
no extrapolation is required. At the
lightest pion mass we find
\begin{equation}
g_A^R = 1.24 \pm 0.04 \,, 
\label{gad}
\end{equation}
in good agreement with the previous determination and the experimental
value. It turns out that 
$g_A^R$ hovers around $\approx 1.1$ for $m_\pi \gtrsim
250\,\mbox{MeV}$, a feature it shares with most other lattice
calculations~\cite{Lin:2011ab}. Only within the last
$100\,\mbox{MeV}$ from the physical point does $g_A^R$ rise to
its final value. This phenomenon is not totally
unexpected, from general arguments~\cite{Jaffe:2001eb} and from
ChEFT~\cite{Khan:2006de,Procura:2006gq,Bernard:2006te}. 
Near the chiral limit ChEFT predicts, following~\cite{Khan:2006de},  
\begin{equation}
\begin{split}
g_A^R(m_\pi) &= g_A^0 - \frac{g_A^{0\;3}}{16 \pi^2 f_0^2}\,m_\pi^2 +
4\left[B_9^r(m_{\pi\,{\rm phys}})-2g_A^0\, 
  B_{20}^r(m_{\pi\,{\rm phys}})\right] m_\pi^2 \\[0.5em]
&-\frac{g_A^{0\;3}+g_A^0/2}{4\pi^2 f_0^2}\,m_\pi^2\,
\ln(m_\pi/m_{\pi\,{\rm phys}}) +
O\left(m_\pi^3\right) \,.
\end{split}
\label{cheft}
\end{equation}
To this order, both sets of chiral
expansions,~\cite{Khan:2006de,Procura:2006gq} 
and~\cite{Bernard:2006te}, are equivalent  
with $B_9=d_{16}$ and $B_{20}=d_{28}$.
In (\ref{cheft}) we have chosen $\lambda =
m_{\pi\,{\rm phys}}$ ($\lambda$ being the scale parameter of the
dimensional regularization). The coupling $B_{9}^r$ cannot be
observed independent of $B_{20}^r$. Taking $B_{20}^r(m_{\pi\,{\rm
    phys}}) \equiv 0$, the preferred value is~\cite{Hemmert:2003cb}   
$\displaystyle B_9^r(m_{\pi\,{\rm phys}}) = (-1.4 \pm
1.2)\,\mbox{GeV}^{-2}$. A fit of the leading order chiral formula
(\ref{cheft}) to the data points in Fig.~\ref{fig4} is shown by the
shaded area. The fit gives $g_A^0 = 1.26(7)$ and
$\displaystyle B_9^r(m_{\pi\,{\rm phys}}) = (-2.1 \pm 
1.0)\,\mbox{GeV}^{-2}$. 

Our results for $f_\pi^R$ are plotted in Fig.~\ref{fig5}. Again, no
extrapolation to the physical point is needed. At the lightest pion
mass we find
\begin{equation}
f_\pi^R = 89.7 \pm 1.5 \pm 1.8\, \mbox{MeV} \,,
\label{fpil}
\end{equation}
using $r_0=0.50(1)\,\mbox{fm}$. The second error in eq.~(\ref{fpil})
is due to the error on $r_0$. 

Instead of taking $f_\pi^R$ at the lowest pion mass,
eq.~(\ref{fpil}), it might be a better idea to include the adjacent
data points in the analysis as well and fit the data by a chiral
ansatz~\cite{Colangelo:2001df},
\begin{equation}
f_\pi^R = f_0 \left[1-\frac{m_\pi^2}{16\pi^2 f_\pi^{R\;2}}
  \ln\left(\Lambda_4^2/m_\pi^2\right)\right]^{-1} + A\,m_\pi^4 \,.
\label{chpt2}
\end{equation}
The result of the fit is shown in Fig.~\ref{fig5}. At the physical
point we obtain 
\begin{equation}
f_\pi^R = 89.6 \pm 1.1\pm 1.8\ \,\mbox{MeV} \,,
\end{equation}
in full agreement with the result (\ref{fpil}).
The main effect is that the statistical error has reduced by
$30\%$. In the chiral limit we obtain $f_0=86(1)\,\mbox{MeV}$, which
agrees with the assumption made in Sec.~\ref{axinf}. A
fit of the chiral ansatz (\ref{chpt2}) to the lowest four data points
with $A=0$ gives the low-energy constant
\begin{equation}
\bar{l}_4 = \ln\left(\Lambda_4^2/m_{\pi\,{\rm phys}}^2\right)= 4.2 \pm
0.1 \,.
\end{equation}

\section{Conclusions}

We have successfully computed the nucleon axial charge and the
pion decay constant in $N_f=2$ lattice QCD with nonperturbatively
$O(a)$ improved Wilson fermions. A novel feature of our calculations is
that we have data at virtually physical pion mass and for a variety of
lattice volumes and spacings at our disposal. While our
simulations at different lattice spacings indicate that our results
are free from discretization effects, simulations on different lattice
volumes indicate the presence of large finite size effects. Two
approaches have been pursued.

The main result of this paper is a determination of $g_A^R$ from the
ratio $g_A/f_\pi$, which is free of finite size effects and
renormalization errors. We found that this ratio has a smooth behavior
as a function of quark mass, and can be essentially described by a
polynomial in $m_\pi^2$, leading to $g_A^R=1.29(5)(3)$ at the physical
point, in excellent agreement with experiment. Here we have used
$r_0=0.50(1)\,\mbox{fm}$ to set the scale, which we obtained from fits to
the nucleon mass~\cite{Bali:2012qs}. This result is in perfect
agreement with ALPHA~\cite{Fritzsch:2012wq}, who finds
$r_0=0.503(10)\,\mbox{fm}$ using $f_K$ to set the 
scale and the same action. In contrast, ETM finds consistently lower
values of $r_0$, $r_0=0.465(6)(14)\,\mbox{fm}$ from the nucleon
mass~\cite{Alexandrou:2009qu} and $r_0=0.420(9)(+10/-11)\,\mbox{fm}$
from using $f_\pi$ to set the scale~\cite{Baron:2009wt}. If correct, this
would raise our number for $g_A$ accordingly.

We attempted a direct calculation of $g_A^R$ and $f_\pi^R$, 
taking account of finite size corrections and
renormalization. To our knowledge, this is the first time finite
size corrections have been applied to $g_A$ at physical pion
masses. Both approaches give consistent results, suggesting that
finite size corrections to both $g_A$ and $f_\pi$ are well described
by ChEFT and ChPT.  


\clearpage
\section*{Appendix: Finite size corrections}

Let us first consider $g_A$. Utilizing
the (nonrelativistic) small scale expansion (SSE) of the ChEFT,
including pion, nucleon (N) and $\Delta(1232)$ degrees of freedom, we
obtain to 
$O(\epsilon^3)$~\cite{Khan:2006de} 
\begin{equation}
\frac{g_A(L)-g_A(\infty)}{g_A(\infty)} = - \frac{m_\pi^2}{4\pi^2 f_0^2}
\sum_{\substack{\vn\\|\vn|\neq 0}} \frac{K_1(\lambda |\vn|)}{\lambda
  |\vn|} + \Delta(L) 
\label{fsga1}
\end{equation}
with 
\begin{equation}
\begin{split}
\Delta(L) &= \frac{g_A^2 m_\pi^2}{6\pi^2 f_0^2}
\sum_{\substack{\vn\\|\vn|\neq 0}} \left[K_0(\lambda
  |\vn|)-\frac{K_1(\lambda |\vn|)}{\lambda |\vn|}\right] \\
          &+ \frac{25 c_A^2 g_1}{81\pi^2 g_A f_0^2} \int_0^\infty
dy\,y\, \sum_{\substack{\vn\\|\vn|\neq 0}} \left[K_0(\lambda(y)
  |\vn|)-\frac{\lambda(y)|\vn|}{3} K_1(\lambda(y) |\vn|)\right] \\
          &- \frac{c_A^2}{\pi^2 f_0^2} \int_0^\infty dy\,y\, 
\sum_{\substack{\vn\\|\vn|\neq 0}} \left[K_0(\lambda(y)
  |\vn|)-\frac{\lambda(y)|\vn|}{3} K_1(\lambda(y) |\vn|)\right] \\
          &+ \frac{8c_A^2m_\pi^2}{27\pi^2 f_0^2\Delta_0} \int_0^\infty
dy\, \sum_{\substack{\vn\\|\vn|\neq 0}}
\left(\frac{\lambda(y)}{\lambda}\right)^2\left[K_0(\lambda(y) 
  |\vn|)-\frac{K_1(\lambda(y) |\vn|)}{\lambda(y) |\vn|}\right] \\
          &- \frac{4c_A^2m_\pi^3}{27\pi f_0^2\Delta_0}  
\sum_{\substack{\vn\\|\vn|\neq 0}} \frac{e^{-\lambda |\vn|}}{\lambda
  |\vn|} \,,
\end{split}
\label{fsga2}
\end{equation}
where $\lambda = m_\pi L$ and $\lambda(y) = f(m_\pi,y) L$ with
$f(m_\pi,y)=\sqrt{m_\pi^2+y^2+2y\Delta_0}$, $\Delta_0$ being the
$\Delta-N$ mass difference. $K_0$ and $K_1$ denote the modified
Bessel functions, and $c_A$ and $g_1$ are the leading axial $\Delta
N$ and $\Delta \Delta$ couplings. The parameter $c_A$ should not be
confused with the improvement coefficient $c_A$ in eq.~(\ref{imp}).

The second term in eq.~(\ref{fsga1}), $\Delta(L)$, receives
contributions from chiral loops, which renormalize the axial charge
and act on intermediate $\Delta$ baryons~\cite{Khan:2006de}. It turns
out that the various contributions to $\Delta(L)$ effectively cancel
each other over a wide range of $\lambda$ values. This has been noticed
by the authors of~\cite{Hall:2012qn} as well. To state an example, let
us consider the $48^3\times 64$ lattice at $\beta=5.29$,
$\kappa=0.13640$. This lattice has the lowest pion mass and is
especially important for our final conclusions. Taking $c_A=1.5$
from~\cite{Gail:2005gz} and 
$g_1=2.16$ from $SU(6)$, we find $-0.044$
for the total contribution, but 
only $+0.001$ for $\Delta(L)$. We thus may assume
\begin{equation}
\frac{g_A(L)-g_A(\infty)}{g_A(\infty)} = - \frac{m_\pi^2}{4\pi^2 f_0^2}
\sum_{\substack{\vn\\|\vn|\neq 0}} \frac{K_1(\lambda |\vn|)}{\lambda
  |\vn|} \,.
\label{fsga3}
\end{equation}

The finite size corrections to $f_\pi$ have been computed
in~\cite{Colangelo:2005gd} within the context of ChPT. To NLO
($\propto m_\pi^2$) the outcome is  
\begin{equation}
\frac{f_\pi(L)-f_\pi(\infty)}{f_\pi(\infty)} = - \frac{m_\pi^2}{4\pi^2
  f_0^2} 
\sum_{\substack{\vn\\|\vn|\neq 0}} \frac{K_1(\lambda |\vn|)}{\lambda
  |\vn|} \,.
\label{fsfpi}
\end{equation}
The NNLO corrections are found to be very small on our configurations
and, thus, can safely be neglected. 

The investigations above show that the leading finite size
corrections to $g_A$ and $f_\pi$, eqs.~(\ref{fsga3}) and
(\ref{fsfpi}), are identical. Once $f_0$, the pion 
decay constant in the chiral limit, has been fixed, expressions
(\ref{fsga3}) and (\ref{fsfpi}) have only one free parameter,
$g_A(\infty)$ and $f_\pi(\infty)$, respectively. 

The NLO correction to the pion mass reads~\cite{Colangelo:2005gd}
\begin{equation}
\frac{m_\pi(L)-m_\pi(\infty)}{m_\pi(\infty)} =  \frac{m_\pi^2}{16\pi^2
  f_0^2} 
\sum_{\substack{\vn\\|\vn|\neq 0}} \frac{K_1(\lambda |\vn|)}{\lambda
  |\vn|} \,.
\label{fsmpi}
\end{equation}
At smaller values of $m_\pi L$, $m_\pi L \lesssim 3$, this expression
alone cannot describe the observed finite size
effects~\cite{Bali:2012qs}. That is not surprising, since in a finite
spatial 
box chiral symmetry does not break down spontaneously. This is because giving
the system enough time it will rotate through all vacua. This results
in a mass gap at vanishing quark
masses~\cite{Leutwyler:1987ak,H&N,Hasenfratz:2009mp}, 
\begin{equation}
m_{\pi\,{\rm res}} = \frac{3}{2 f_0^2 L^3 (1+\Delta)}
\label{mres}
\vspace*{-0.1cm}
\end{equation}
with
\begin{equation}
\begin{split}
\Delta &= \frac{2}{f_0^2 L^2}\,0.2257849591 \\
       &+ \frac{1}{f_0^4 L^4}\,
\left[0.088431628 - \frac{0.8375369106}{3\pi^2}\,\Big(\frac{1}{4}
  \ln{\left(\Lambda_1^2 L^2\right)} +  \ln{\left(\Lambda_2^2
    L^2\right)}\Big)\right]\, , 
\end{split}
\end{equation}
where $\Lambda_i$ are the intrinsic scale parameters of the low-energy
constants $\bar{l}_i = \ln\left(\Lambda_i^2/m_{\pi\,{\rm
  phys}}^2\right)$~\cite{Colangelo:2001df}, with $m_{\pi\,{\rm phys}}$ being the
physical pion mass. In~\cite{Bietenholz:2010az} we
found that the pion mass extrapolates indeed to a finite value 
in the chiral limit, in good agreement with the expected result
(\ref{mres}). This also has an effect on $m_\pi$ in the region of small,
but nonvanishing, quark masses~\cite{Bietenholz:2010az}. We
thus expect the finite size correction to be effectively given by 
\begin{equation}
m_\pi(L)=m_\pi(\infty)+ \frac{m_\pi^3}{16\pi^2 f_0^2} 
\sum_{\substack{\vn\\|\vn|\neq 0}} \frac{K_1(\lambda |\vn|)}{\lambda
  |\vn|} + \frac{3\, c(m_\pi)}{2 f_0^2 L^3 (1+\Delta)}
\label{com}
\end{equation}
with the parameter $c(m_\pi)$ rapidly dropping to zero at larger pion masses.

\section*{Acknowledgments}

The gauge configurations were generated using the BQCD
code~\cite{Nakamura:2010qh} on  
the BlueGene/L and BlueGene/P at NIC (J\"ulich), the BlueGene/L at 
EPCC (Edinburgh), the SGI ICE 8200 at HLRN (Berlin and Hannover), and
on QPACE. The Chroma software library~\cite{Edwards:2004sx} was used
in the data analysis. This work would not have been possible without
the input of Dirk Pleiter. We thank him most sincerely for his
contributions. Benjamin
Gl\"a{\ss}le computed some 
two- and three-point functions for us, which we gratefully
acknowledge. This work has been supported partly 
by the EU grants 283286 (HadronPhysics3) and 227431 (HadronPhysics2),
and by the DFG under contract SFB/TR 55 (Hadron Physics from Lattice
QCD). JMZ is supported by the Australian Research Council grant
FT100100005. We thank all funding agencies.


\begin{thebibliography}{99}

\bibitem{Edwards:2005ym} 
  R.~G.~Edwards, G.~T.~Fleming, Ph.~H\"agler, J.~W.~Negele,
  K.~Orginos, A.~V.~Pochinsky, D.~B.~Renner, D.~G.~Richards and
  W.~Schroers, 
  Phys.\ Rev.\ Lett.\  {\bf 96}, 052001 (2006)
  [hep-lat/0510062].

\bibitem{Yamazaki:2008py} 
  T.~Yamazaki, Y.~Aoki, T.~Blum, H.-W.~Lin, M.-F.~Lin, S.~Ohta,
  S.~Sasaki, R.~J.~Tweedie and J.~M.~Zanotti, 
  Phys.\ Rev.\ Lett.\  {\bf 100}, 171602 (2008)
  [arXiv:0801.4016 [hep-lat]].

\bibitem{Alexandrou:2010hf} 
  C.~Alexandrou, M.~Brinet, J.~Carbonell, M.~Constantinou,
  P.~A.~Harraud, P.~Guichon, K.~Jansen, T.~Korzec and M. Papinutto,
  Phys.\ Rev.\ D {\bf 83}, 045010 (2011)
  [arXiv:1012.0857 [hep-lat]].

\bibitem{Capitani:2012gj} 
  S.~Capitani, M.~Della Morte, G.~von Hippel, B.~J\"ager,
  A.~J\"uttner, B.~Knippschild, H.~B.~Meyer and H.~Wittig, 
  Phys.\ Rev.\ D {\bf 86}, 074502 (2012)
  [arXiv:1205.0180 [hep-lat]].

\bibitem{Lin:2011ab} 
  For reviews see: H.-W.~Lin, arXiv:1112.2435 [hep-lat];
  H.-W.~Lin, 
  PoS LATTICE {\bf 2012}, 013 (2012)
  [arXiv:1212.6849 [hep-lat]].

\bibitem{Martinelli:1994ty} 
  G.~Martinelli, C.~Pittori, C.~T.~Sachrajda, M.~Testa and A.~Vladikas,
  Nucl.\ Phys.\ B {\bf 445}, 81 (1995)
  [hep-lat/9411010];\\
  M.~G\"ockeler, R.~Horsley, H.~Oelrich, H.~Perlt, D.~Petters,
  P.~E.~L.~Rakow, A.~Sch\"afer, G.~Schierholz and A.~Schiller, 
  Nucl.\ Phys.\ B {\bf 544}, 699 (1999)
  [hep-lat/9807044].

\bibitem{Gockeler:2010yr} 
  M.~G\"ockeler, R.~Horsley, Y.~Nakamura, H.~Perlt, D.~Pleiter,
  P.~E.~L.~Rakow, A.~Sch\"afer, G.~Schierholz, A.~Schiller,
  H.~St\"uben and J.~M.~Zanotti, 
  Phys.\ Rev.\ D {\bf 82}, 114511 (2010)
  [arXiv:1003.5756 [hep-lat]].

\bibitem{Constantinou:2012dt} 
  M.~Constantinou, M.~Costa, M.~G\"ockeler, R.~Horsley,
  H.~Panagopoulos, H.~Perlt, P.~E.~L.~Rakow, G.~Schierholz and A.~Schiller, 
  PoS LATTICE {\bf 2012}, 239 (2012)
  [arXiv:1210.7737 [hep-lat]].

\bibitem{Beane:2004rf} 
  S.~R.~Beane and M.~J.~Savage,
  Phys.\ Rev.\ D {\bf 70}, 074029 (2004)
  [hep-ph/0404131].

\bibitem{Khan:2006de} 
  A.~Ali~Khan, M.~G\"ockeler, P.~H\"agler, T.~R.~Hemmert, R.~Horsley,
  D.~Pleiter, P.~E.~L.~Rakow, A.~Sch\"afer, G.~Schierholz,
  T.~Wollenweber and J.~M.~Zanotti,
  Phys.\ Rev.\ D {\bf 74}, 094508 (2006)
  [hep-lat/0603028].

\bibitem{Colangelo:2005gd} 
  G.~Colangelo, S.~D\"urr and C.~Haefeli,
  Nucl.\ Phys.\ B {\bf 721}, 136 (2005)
  [hep-lat/0503014].

\bibitem{Pleiter:2011gw} 
  S.~Collins, M.~G\"ockeler, Ph.~H\"agler, T.~Hemmert, R.~Horsley,
  Y.~Nakamura, A.~Nobile, H.~Perlt, D.~Pleiter, P.~E.~L.~Rakow,
  A.~Sch\"afer, G.~Schierholz, A.~Sternbeck, H.~St\"uben, F.~Winter
  and J.~M.~ Zanotti, 
  PoS LATTICE {\bf 2010}, 153 (2010)
  [arXiv:1101.2326 [hep-lat]].

\bibitem{Bali:2012qs} 
  G.~S.~Bali, P.~C.~Bruns, S.~Collins, M.~Deka, B.~Gl\"a{\ss}le,
    M.~G\"ockeler, L.~Greil, T.~R.~Hemmert, R. Horsley, J. Najjar,
    Y. Nakamura, A. Nobile, D. Pleiter, P.~E.~L. Rakow, A. Sch\"afer,
    R. Schiel, G. Schierholz, A. Sternbeck and J.~M.~ Zanotti,
  Nucl.\ Phys.\ B {\bf 866}, 1 (2013)
  [arXiv:1206.7034 [hep-lat]].

\bibitem{Della Morte:2005se} 
  M.~Della Morte, R.~Hoffmann and R.~Sommer,
  JHEP {\bf 0503}, 029 (2005)
  [hep-lat/0503003].

\bibitem{Gockeler:1995wg} 
  M.~G\"ockeler, R.~Horsley, E.~-M.~Ilgenfritz, H.~Perlt,
  P.~E.~L.~Rakow, G.~Schierholz and A.~Schiller, 
  Phys.\ Rev.\ D {\bf 53}, 2317 (1996)
  [hep-lat/9508004].

\bibitem{Capitani:1998ff} 
  S.~Capitani, M.~G\"ockeler, R.~Horsley, B.~Klaus, H.~Oelrich,
  H.~Perlt, D.~Petters, D.~Pleiter, P.~E.~L.~Rakow, G.~Schierholz,
  A.~Schiller and P. Stephenson,  
  Nucl.\ Phys.\ Proc.\ Suppl.\  {\bf 73}, 294 (1999)
  [hep-lat/9809172].

\bibitem{Owen:2012ts} 
  B.~J.~Owen, J.~Dragos, W.~Kamleh, D.~B.~Leinweber, M.~S.~Mahbub,
  B.~J.~Menadue and J.~M.~Zanotti, 
  arXiv:1212.4668 [hep-lat].

\bibitem{Gockeler:2005mh} 
  M.~G\"ockeler, R.~Horsley, D.~Pleiter, P.~E.~L.~Rakow, G.~Schierholz,
  W.~Schroers, H.~St\"uben and J.~M.~Zanotti, 
  PoS LAT {\bf 2005}, 063 (2006)
  [hep-lat/0509196].

\bibitem{Colangelo:2001df}
  G.~Colangelo, J.~Gasser and H.~Leutwyler,
  Nucl.\ Phys.\  B {\bf 603}, 125 (2001)
  [arXiv:hep-ph/0103088].

\bibitem{Procura:2006gq} 
  M.~Procura, B.~U.~Musch, T.~R.~Hemmert and W.~Weise,
  Phys.\ Rev.\ D {\bf 75}, 014503 (2007)
  [hep-lat/0610105].

\bibitem{Colangelo:2003hf} 
  G.~Colangelo and S.~D\"urr,
  Eur.\ Phys.\ J.\ C {\bf 33}, 543 (2004)
  [hep-lat/0311023].

\bibitem{Bietenholz:2010az} 
  W.~Bietenholz, M.~G\"ockeler, R.~Horsley, Y.~Nakamura, D.~Pleiter,
  P.~E.~L.~Rakow, G.~Schierholz and J.~M.~Zanotti, 
  Phys.\ Lett.\ B {\bf 687}, 410 (2010)
  [arXiv:1002.1696 [hep-lat]].

\bibitem{Sint:1997jx} 
  S.~Sint and P.~Weisz,
  Nucl.\ Phys.\ B {\bf 502}, 251 (1997)
  [hep-lat/9704001].

\bibitem{Jaffe:2001eb} 
  R.~L.~Jaffe,
  Phys.\ Lett.\ B {\bf 529}, 105 (2002)
  [hep-ph/0108015].

\bibitem{Bernard:2006te}
  V.~Bernard and U.~-G.~Meissner,
  Phys.\ Lett.\ B {\bf 639}, 278 (2006)
  [hep-lat/0605010].

\bibitem{Hemmert:2003cb} 
  T.~R.~Hemmert, M.~Procura and W.~Weise,
  Phys.\ Rev.\ D {\bf 68}, 075009 (2003)
  [hep-lat/0303002].

\bibitem{Hall:2012qn} 
  N.~L.~Hall, A.~W.~Thomas, R.~D.~Young and J.~M.~Zanotti,
  arXiv:1205.1608 [hep-lat].

\bibitem{Fritzsch:2012wq} 
  P.~Fritzsch, F.~Knechtli, B.~Leder, M.~Marinkovic, S.~Schaefer,
  R.~Sommer, and F.~Virotta
  Nucl.\ Phys.\ B {\bf 865}, 397 (2012)
  [arXiv:1205.5380 [hep-lat]].

\bibitem{Alexandrou:2009qu} 
  C.~Alexandrou, R.~Baron, J.~Carbonell, V.~Drach, P.~Guichon,
  K.~Jansen, T.~Korzec and O.~P\`ene,
  Phys.\ Rev.\ D {\bf 80}, 114503 (2009)
  [arXiv:0910.2419 [hep-lat]].

\bibitem{Baron:2009wt} 
  R.~Baron, Ph.~Boucaud, P.~Dimopoulos, F.~Farchioni, R.~Frezzotti,
  V.~Gimenez, G.~Herdoiza, K.~Jansen, V.~Lubicz, C.~Michael,
  G.~M\"unster, D.~Palao, G.~C.~Rossi, L.~Scorzato, A.~Shindler,
  S.~Simula, T.~Sudmann, C.~Urbach and U.~Wenger,
  JHEP {\bf 1008}, 097 (2010)
  [arXiv:0911.5061 [hep-lat]].

\bibitem{Gail:2005gz} 
  T.~A.~Gail and T.~R.~Hemmert,
  Eur.\ Phys.\ J.\ A {\bf 28}, 91 (2006)
  [nucl-th/0512082].

\bibitem{Leutwyler:1987ak}
  H.~Leutwyler,
  Phys.\ Lett.\  B {\bf 189}, 197 (1987).

\bibitem{H&N}
  P.~Hasenfratz and F.~Niedermayer, Z.\ Phys.\ B {\bf 92}, 91 (1993)
  [arXiv:hep-lat/9212022].

\bibitem{Hasenfratz:2009mp}
  P.~Hasenfratz,
  Nucl.\ Phys.\  B {\bf 828}, 201 (2010)
  [arXiv:0909.3419 [hep-th]].

\bibitem{Nakamura:2010qh} 
  Y.~Nakamura and H.~St\"uben,
  PoS LATTICE {\bf 2010}, 040 (2010)
  [arXiv:1011.0199 [hep-lat]].

\bibitem{Edwards:2004sx} 
  R.~G.~Edwards and B.~Jo{\'o},
  Nucl.\ Phys.\ Proc.\ Suppl.\  {\bf 140}, 832 (2005)
  [hep-lat/0409003].

\end{thebibliography}
\end{document}